# Ultrashort intense-field optical vortices produced with laser-etched mirrors

**J. Strohaber, T. Scarborough, and C. J. G. J. Uiterwaal**

*Department of Physics & Astronomy, The University of Nebraska- Lincoln,
Behlen Lab – City Campus, Lincoln, NE 68588-0111*
*jstroha1@bigred.unl.edu , scar603@bigred.unl.edu , cuiterwaal2@unl.edu*

**Abstract:** We introduce a simple and practical method to create ultrashort intense optical vortices for applications involving high-intensity lasers. Our method utilizes femtosecond laser pulses to laser-etch grating lines into laser-quality gold mirrors. These grating lines holographically encode an optical vortex. We derive mathematical equations for each individual grating line to be etched, for any desired (integer) topological charge. We investigate the smoothness of the etched grooves. We show that they are smooth enough to produce optical vortices with an intensity that is only a few percent lower than in the ideal case. We demonstrate that the etched gratings can be used in a folded version of our 2*f*-2*f* setup [Mariyenko *et al.*, Opt. Express **19,** 7599 (2005)] to compensate angular dispersion. Lastly, we show that the etched gratings withstand intensities of up to $10^{12}$ W/cm$^2$.

## 1. Introduction

Since the development of mode locking and chirped-pulsed amplification, intense and ultrashort laser pulses have been ubiquitously used in laboratories to study laser-matter interactions under extreme conditions[1,2]. What makes intense, ultrashort pulses so interesting is that peak intensities of $10^{15}$ W/cm$^2$ and more can be achieved. The corresponding electric fields result in a force on an atomic electron that is comparable to the force the nucleus exerts on it. Thus the pulse is capable of liberating an electron from its parent nucleus. Much research has been carried out in this area, known as intense-field ionization; some highlights are multiphoton ionization, tunneling ionization, and above threshold ionization[1,2].

Another field, evolved from research in wavefield dislocations[3] and laser modes within cavities[4], is known as singular optics[5]. An archetypical example of phase singularities in optics is the Laguerre-Gaussian[6] tranverse paraxial beam mode $\text{LG}_{p=0}^{m=1}$, also known as the 'donut mode' (here $m$ is the azimuthal mode number and $p$ is the radial mode number)[5,6]. LG modes have an azimuthal phase dependency of $\exp(im\theta)$ ($\theta$ is the azimuthal angle). This phase dependency causes the electric field to be undetermined on the optical axis[7]. Consequently, the field amplitude vanishes there; it is the location of an optical vortex (OV) with topological charge $m$[5]. Ince-Gaussian modes[8,9] provide a connection between the Hermite-Gaussian[6] and LG modes. Helical Ince-Gaussian (HIG) modes possess a number of vortices on a straight line. These modes were experimentally realized by symmetry-breaking of a laser cavity[10] and later also produced holographically[11]. The LG and helical IG beams carry optical orbital angular momentum (OAM). Currently, in high-field physics there have been no experiments performed to investigate the effects of this quantity on atomic or molecular systems. Because we are interested in investigating possible effects of OAM in ionization processes, this paper is devoted to presenting a simple method to produce OAM-containg beams (in particular, OVs) by laser-etching gratings lines into laser-quality mirrors.

Different methods have been devised to experimentally realize OVs in the laboratory. Among these methods are: spiral phase plates (SPP)[12,13,14], and methods based on computer generated holograms (CGH)[5,6], either using photographically produced gratings[15] or, more recently, spatial light modulators (SLM)[11,16]. Each of these methods has its advantages and disadvantages depending on the application. For example, photographic methods offer high resolution capabilities compared to an SLM or an SPP[13]. However, photographic and SPP methods are inflexible compared to SLMs (which can be easily reprogrammed). Holograms recorded on photographic film are not suitable for high intensities as the film will be damaged. SLMs are more robust than film but are still less suitable for high intensities than SPPs[12,13].

When producing intense, ultrashort optical vortices, some considerations need to be taken into account to obtain *azimuthally pure beams,* i.e. beams without mode impurities that depend on the azimuthal mode number $m$, since this determines the beam's OAM, the quantity of our physical interest. Both on-axis (Gabor) holography and off-axis (Leith and Upatnieks) holography[17] have been used to produce OVs. For on-axis holography, SPPs have been successfully used[5]. For off-axis holography, gratings are typically used[5,15,16,18]. Upon illuminating either of these elements with monochromatic light having a Gaussian intensity profile we obtain OVs that are generally a superposition of pure LG states[5,6]. The contributing LG modes have the same azimuthal mode number $m$ but different radial mode numbers $p$. The two elements differ in that spiral phase plates are near 100% power efficient[18], while gratings are typically less efficient[13,16,19].

For broadband radiation (e.g. ultrashort pulses), spiral phase plates are not suitable[13,19]. These plates are designed for a specific wavelength, so that for all other spectral components in the pulse there is a mismatch of the step height. This results in mode impurities which depend upon the azimuthal mode number[20]. For gratings, the wavelength dependence is manifest in different spectral components emerging from the grating at different angles (angular chirp)[16,21]. The azimuthal mode impurities introduced by an SPP result in an intensity profile that is not cylindrically symmetric. In nonlinear laser-matter interactions such as multiphoton ionization, the use of azimuthally impure beams could cause difficulties when comparing integrated ion yields to theories. No method has been reported to correct the azimuthal mode impurities an SPP causes. In the case of OVs produced with gratings, wavelength dependent angular dispersion can be compensated, but at the expense of power[13,15,16,19]. Thus, it is desirable to have efficient OV gratings that can withstand high intensities. In this Letter, we describe a method to produce such gratings using laser-etching of gold-coated mirrors.

This text is organized as follows. In Sec. 2 we derive the mathematical formulas that describe the grating lines to be etched ('skeleton equations'). In Sec. 3 we give details of the fabrication process. In Sec. 4 we discuss the performance of the laser-etched gratings. Finally, we draw conclusions.

## 2. Skeleton Equations for the Grating Lines

To manufacture our gratings, we selectively removed the gold plating from laser-quality gold mirrors by focusing femtosecond laser radiation onto their surface. Throughout this paper we will refer to these laser-etched mirrors as "LG-mirrors." In this section we provide a prescription of the so-called skeleton equations[7] that we used to laser-etch grating lines into the blank mirrors with the help of MATLAB and LabVIEW codes.

It is known that the phase-only interference of an optical vortex $\exp[i(k_z z - m\theta + \Phi_0)]$ with that of an oblique plane reference wave $\exp[i(k_z z + k_x x)]$ results in an interference pattern that gives the grating structure needed to produce an optical vortex beam[5,7]:

$$I(x,y) = \frac{1}{2}\left\{1 + \cos\left[m\arctan\left(\frac{y}{x}\right) - 2\pi K x + \Phi_0\right]\right\}. \tag{1}$$

Here $m = 0, \pm 1, \pm 2, \pm 3, \cdots$ is the topological charge of the OV to be produced, $K \equiv k_x / 2\pi$ is the number of grating lines per unit length. and $\Phi_0$ is an arbitrary phase factor. Relationships like Eq. (1) are commonly used in the production of gratings. For instance, for a binary grating we take the value 1 when the interference term is positive, and the value zero elsewhere[7]:

$$T(x, y) = \begin{cases} 0 & \cos(\Phi) \leq 0 \\ 1 & \cos(\Phi) > 0 \end{cases} \quad (2)$$

Here $\Phi = m \arctan(y/x) - 2\pi Kx + \Phi_0$. The prescription in Eq. (2) gives binary gratings like the ones shown in Fig. 1.

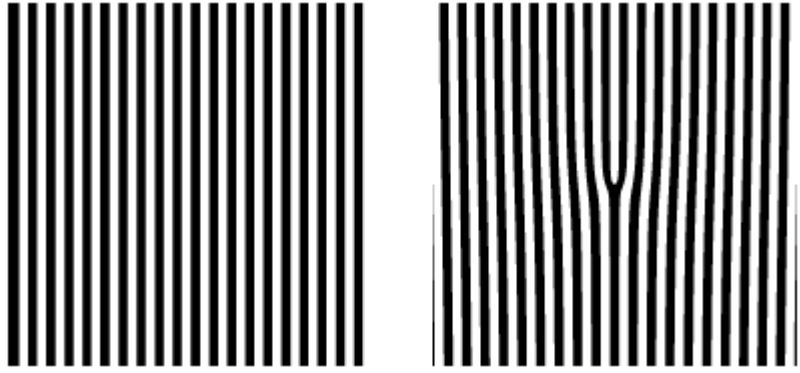

**Fig. 1.** Computer-generated binary grating patterns based on Eq. 2, for $\Phi_0 = -\frac{1}{2}\pi$. Each pattern consists of 1000 × 1000 pixels, where pixels for which $T = 0$ are rendered black and $T = 1$ white. The pattern on the left has no encoded vortex ($m = 0$). The pattern on the right has a single fringe bifurcation ($m = 1$); when used as a hologram it gives rise to vortices with topological charge ±1 in the ± first diffraction order.

For our laser-etching fabrication of holographic gratings we use Eq. (1) to find where each groove needs to be etched. For this purpose we set the interference term in Eq. (1) equal to an arbitrary constant, $c$ (between −1 and +1), which gives the skeleton equations. Setting $\Phi_0 = 0$ for simplicity and choosing $c = 1$ we obtain

$$m \arctan\left(\frac{y}{x}\right) = 2\pi Kx + \arccos(c) \;[\text{mod } 2\pi] = 2\pi(n + Kx). \quad (3)$$

The periodicity of the cosine function in Eq. (1) gives rise to the modulo term $2\pi n$ in Eq. (3), where $n = 0, \pm 1, \pm 2, \cdots$. This integer has a unique value for each grating line, so we call it the grating line number. If, in Eq. (3), we choose $m = 0$, the skeleton equations reduce to that of a straight-line grating $x = -n/K$, as shown in Fig. 2(a). For nonzero (but integer) $m$ we solve Eq. (3) for $y$ to obtain the skeleton equations for the OV holograms:

$$y(x) = x \tan\left[\frac{2\pi}{m}(n + Kx)\right] \quad (4)$$

The periodicity of the tangent function in Eq. (4) gives unwanted branches in the solutions. We exclude these branches by setting conditions on the angular argument of the tangent function for three different regions of the grating plane: positive $x$ half-plane, negative $x$ half-plane, and the line $x = 0$. This latter line case will be discussed later. For the skeleton equations in the regions $x > 0$ and $x < 0$, we impose the following angular conditions ($\theta =$ azimuthal angle):

$$\begin{aligned} x > 0 &\Rightarrow -\tfrac{1}{2}\pi < \theta < \tfrac{1}{2}\pi \\ x < 0 &\Rightarrow \tfrac{1}{2}\pi < \theta < \tfrac{3}{2}\pi \end{aligned} \quad (5)$$

Substituting the argument of the tangent function $\theta = 2\pi(n + Kx)/m$ in the inequalities of Eq. (5) we find the following conditions on the skeleton equations:

$$\begin{aligned} \text{For } x > 0: \quad & -\frac{1}{K}\left(\frac{m}{4}+n\right) \leq x \leq \frac{1}{K}\left(\frac{m}{4}-n\right) \text{ and } n < \frac{m}{4}; \\ \text{for } x < 0: \quad & \frac{1}{K}\left(\frac{m}{4}-n\right) \leq x \leq \frac{1}{K}\left(\frac{3m}{4}-n\right) \text{ and } n > \frac{m}{4}. \end{aligned} \quad (6)$$

The last set of inequalities ($n < m/4$ and $n > m/4$) give the integer grating line numbers for which the first set of inequalities for $x$ hold. For the equation of the line $x = 0$, the argument of the tangent function is degenerate, *viz*.

$$\theta = \begin{cases} \tfrac{1}{2}\pi & y > 0 \\ \text{undefined} & y = 0 \\ -\tfrac{1}{2}\pi & y < 0 \end{cases} \quad (7)$$

Again using $\theta = 2\pi(n + Kx)/m$, we rewrite Eq. (7) as

$$n = \begin{cases} \dfrac{m}{4} & y > 0 \\ \text{undefined} & y = 0 \\ -\dfrac{m}{4} & y < 0 \end{cases} \quad (8)$$

Since $n$ is an integer $n = 0, \pm 1, \pm 2, \ldots$, Eq. (8) can only apply when $m = 0, \pm 4, \pm 8, \ldots$, and the straight line $x = 0$ only needs to be considered for these $m$-values. Note that if a value other than zero was chosen for $\Phi_0$, there could be a line in the positive $y$ half plane and no line in the negative $y$ half plane or vice versa. Summarizing, a grating that produces an OV beam can be drawn one line at a time by using Eq. (4) (skeleton equations), Eq. (6) (which places limits on the $x$-values), and Eq. (8) (for the center line, when needed). To verify that the equations are correct we drew the patterns seen in Fig. 2. We used the same line-by-line drawing approach to laser-etch our mirror holograms—one groove at a time (see next section).

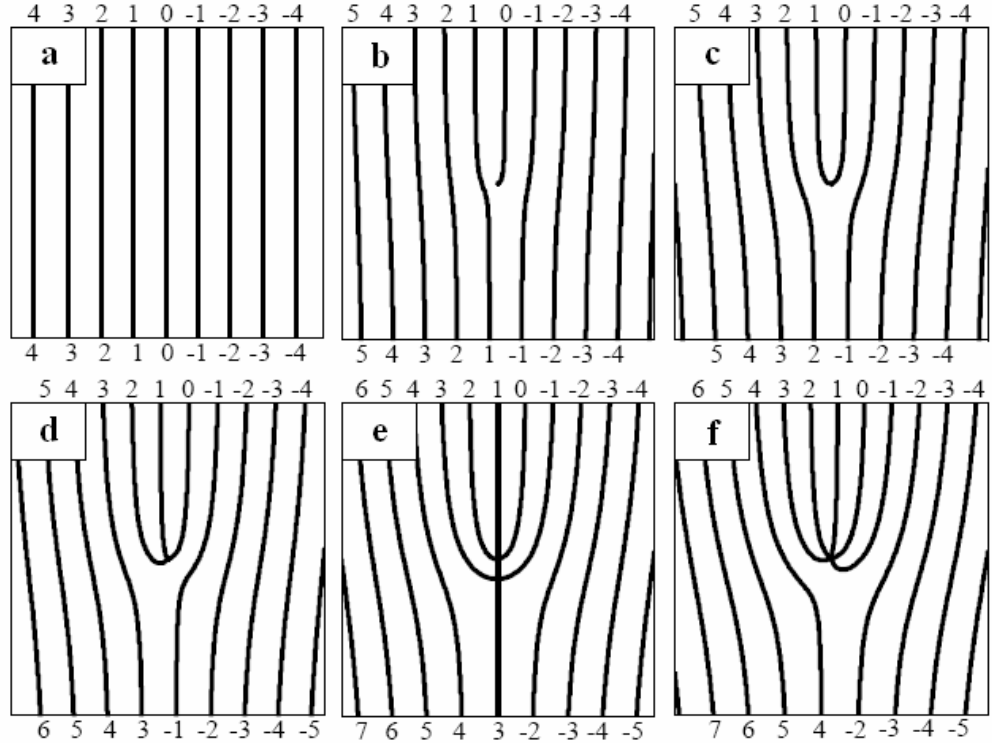

**Fig. 2.** Binary (line) grating holograms constructed using Eq. (4), Eq. (6), and Eq. (8). The holograms are for a topological charge $m$ equal to 0 (a), 1 (b), 2 (c), 3 (d), 4 (e), and 5 (f), all for $\Phi_0 = 0$. Labeled on each hologram are the grating line numbers $n$ for each line.

## 3. Grating Fabrication and Grating Quality

To laser-etch the grating lines into the gold mirrors, we first built a motorized X-Y translation stage. Two unmotorized translation stages (Standa, sensitivity 1 μm, maximum travel 150 mm) were mounted perpendicular to each other as shown in Fig. 3. The resulting X-Y stage was used to move the mirror relative to the fixed focus of the laser beam. Stepper motors (Arrick Robotics, angular step size of 1.8 degrees) were connected to the knobs of each of the translation stages via stainless steel bellow couplers. These couplers compensated for small misalignments between the stepper motors and the translation stages. The stepper motors were connected to an Arrick Robotics MD-2 dual stepper motor driver which we controlled by a laboratory PC. Experimental tests showed that translations of 5 μm could be reliably reproduced. A schematic diagram of the complete setup is shown in Fig. 3.

Laser radiation from a Spectra-Physics Spitfire Ti:sapphire laser having pulse durations of ~50 fs, center wavelength of 800 nm and a maximum output power of ~2.3 W was used in the etching process. The laser beam was focused with an achromatic microscope objective having a magnifying power of 4.0×, numerical aperture of 0.2, and a focal length of 30.8 mm. The objective was independently mounted to a manually controlled vertical translation stage above the motorized X-Y translation stage. The manual translation stage allowed for micrometer positioning of the focus onto the mirror surface.

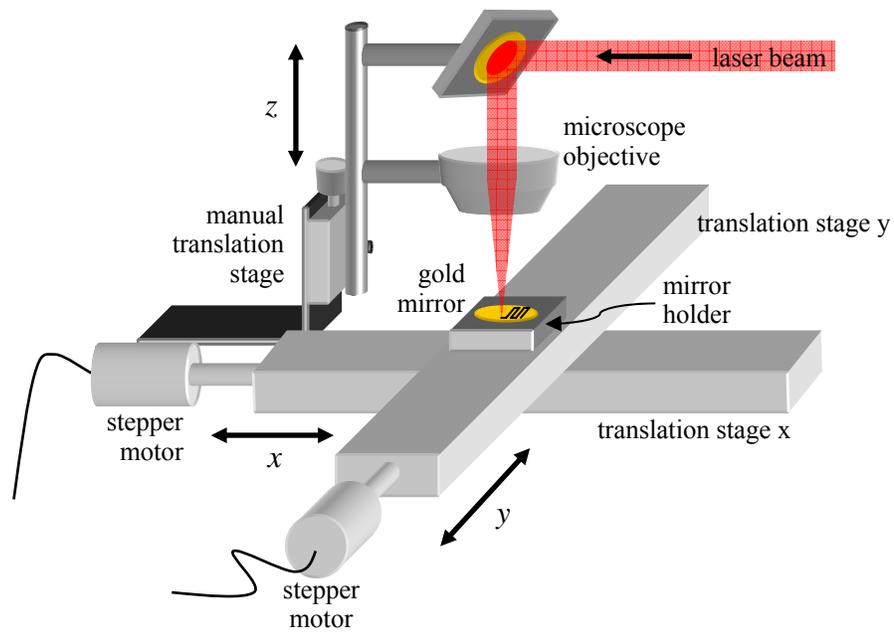

**Fig. 3.** Schematic of the setup used for laser-etching the mirrors. The laser beam is shown entering the setup from the top right. A mirror mounted at 45° with respect to the incoming beam sends the beam through an achromatic microscope objective with a focal length of 30.8 mm. The objective was mounted to a manual vertically-positioned translation stage having a micron resolution. The beam is focused onto a mirror which is mounted on a motorized X-Y translation stage controlled by stepper motors. The resulting resolution in the translation is 5 μm in both the X and Y directions.

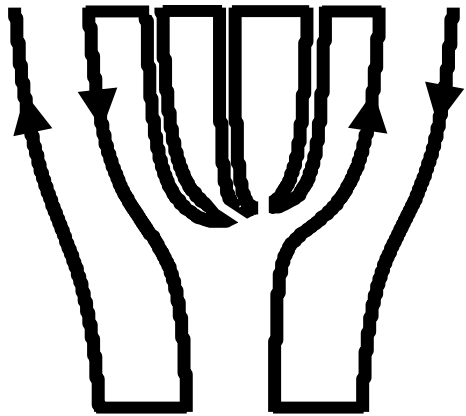

**Fig. 4.** Diagram displaying the method used to etch the gratings as a single continuous line. Arrows show the direction of motion of the focused laser beam relative to the mirror. For clarity, the center lines in this $m = 3$ grating are shown separated. Each of these center lines was etched down to the center of the grating and back up the exact path.

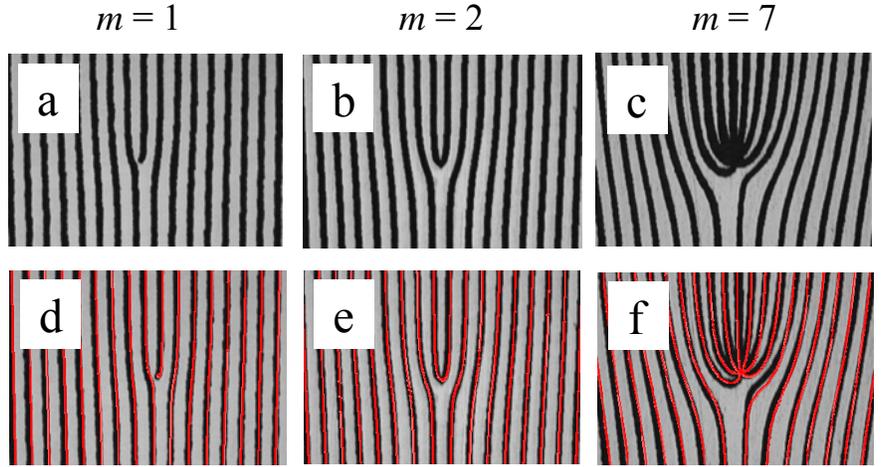

**Fig. 5.** Grey-scale microscopic images of the LG mirrors with 20× magnification for gratings with topological charge (top row, left to right) 1, 2, and 7. The grey portions represent the reflective gold coating, while the black lines are the regions where the gold has been removed by the focused femtosecond radiation. The bottom row shows the same images, overlayed with the skeleton lines (red) calculated from Eq.(3).

We used MATLAB code along with the line equations and their limits discussed in Sec. 2 to create two matrices, one encoding for *x*-translations and the other encoding for *y*-translations. Each column of the matrices was labeled by individual grating lines. These matrices were digitized into steps of 5 µm. LabVIEW code was used to interpret these matrices and subsequently control the movement of the translation stages.

The laser radiation was attenuated to between 60–120 mW before focusing. We found through simple trial-and-error that these powers yielded groove widths that were approximately 1/2 a grating period. This ratio was chosen because for binary amplitude gratings the first order diffraction efficiency $\eta(R) = \sin(\pi R)/\pi$ [17,19] (where $R = Kd$ is the ratio of line width to grating constant) is maximum for $R = 1/2$. Intensities in this power range were high enough to exceed the damage threshold of the gold and remove the reflective coating on the mirrors. The grating lines were etched as one continuous path, as shown in Fig. 4. The resulting laser-etched grating patterns are shown in Figs. 5(a,b,c) for $m = 1, 2$, and $7$. In Figs. 5(d,e,f) the same etched patterns are shown again, together with the calculated lines from the skeleton equations (in red). The etched lines agree with these calculated lines.

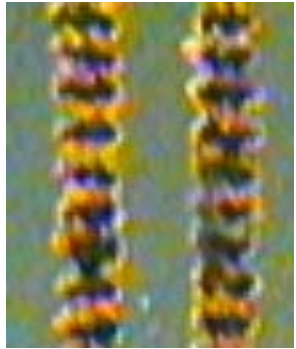

**Fig. 6.** When the laser focus is moved too fast we obtain a trail of separate pits instead of the desired continuous groove, as this microscopic image shows. Here, pits burnt by individual laser pulses no longer spatially overlap. The size of the pits is approximately 20 µm.

As we expected, we had to move the laser focus sufficiently slow over the gold surface to etch a continuous groove with our pulsed laser, which has a repetition rate of 1 kHz. If we set the speed too large we obtained a useless series of pits instead of a groove, as shown in Fig. 6, with each pit burnt by a single pulse of the laser. To investigate the smoothness of our continuous grooves we recorded and analyzed close-up microscopic images of the gratings. Figure 7 gives a typical impression of the detailed groove shape. We investigated how the quality of the groove shape (its smoothness) depends on how fast the focus is moved over the surface to be etched, and also how it is affected by the power used.

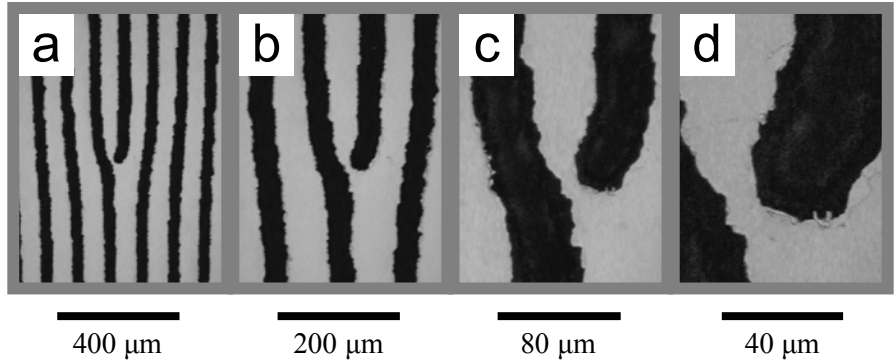

| 400 μm | 200 μm | 80 μm | 40 μm |

**Fig. 7.** Images at various magnifications of an $m=1$ LG-mirror produced by laser etching a laser-quality gold mirror (grey is the gold plating and black is the groove). The grating constant is about 100 μm. The images give an impression of the quality in which our setup can currently laser-etch grating lines.

To investigate the power dependence, we etched a mirror with ~100 mW, and another with only ~60 mW. For a representative groove segment of each mirror we then determined the spread in groove width. Using an edge detection routine in MATLAB, we identified the groove edges (red curves in Figs. 8(a,c)), and we then sampled the groove width at regular intervals. The histograms in Figs. 8(b,d) show the distributions of groove widths we found. We define the groove smoothness as $S = 1 - \Delta d / d$ where $\Delta d$ is the full-width at half-maximum of Gaussian fits to the histograms of Fig. 8, and $d$ is the average groove width. The groove for 100 mW (Fig. 8(a)) shows a larger spread in line width, ($d = 34.9$ μm, $\Delta d = 4.3$ μm, $S = 88$ %), than the one for 60 mW ($d = 41.5$ μm, $\Delta d = 2.4$ μm, $S = 94$ %). The difference in smoothness resulted from decreasing the laser power. We concluded that for 100 mW too much laser power was being delivered to the mirror to properly laser-etch. This was also apparent in the debris-field on the surface of the mirror after the etching process. By contrast, the 60-mW mirror had very little debris on its surface. In either case, the groove smoothness appears acceptable. MATLAB simulations showed that insufficient smoothness causes a noise background in the far-field, but, for our case, only at an estimated intensity level of $-40$ dB to $-30$ dB. This is accompanied by a loss of intensity in the vortex-beam of no more than 5% of the ideal vortex intensity.

## 4. Performance of the Laser-Etched Gratings

We used several optical setups to experimentally demonstrate mode quality, angular dispersion compensation, and the maximum intensity the LG-mirrors can withstand. In our first experiment we used a Michelson interferometer (Fig. 9) to observe the resulting intensity patterns and phases of the optical vortices produced from $m=1$, $m=2$, and $m=7$ LG-mirrors. A removable diverging lens, L, was placed in the reference arm which produces a reference beam having either a plane or spherical wavefront. The reference mirror $M_{ref}$ was mounted onto a translation stage to adjust for differences in optical path length between the two arms of the interferometer. A 50/50 beam splitter was used to split the incoming femtosecond radiation.

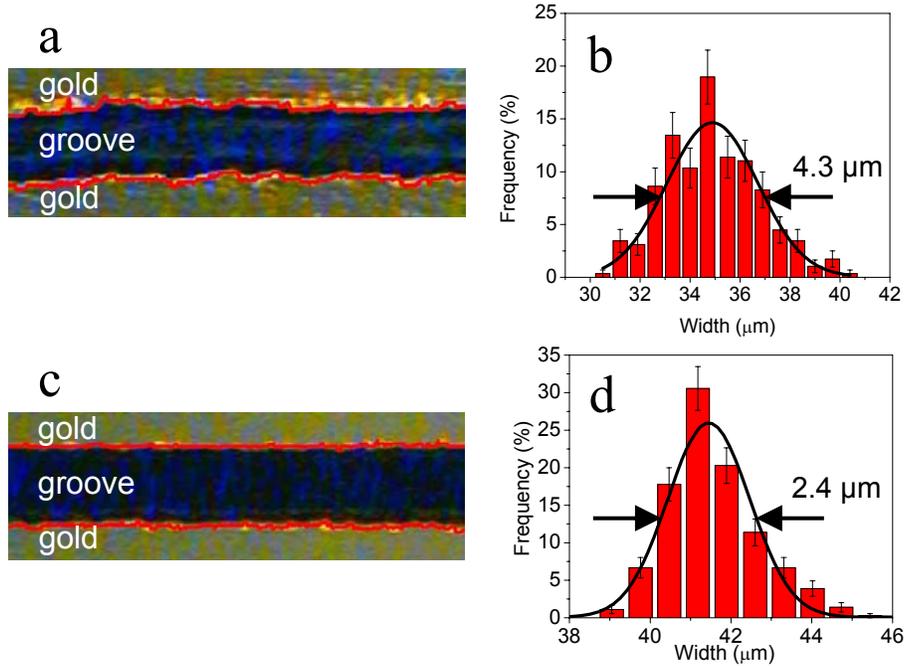

**Fig. 8.** Analysis of groove smoothness: Close-up images of LG-mirrors etched with an average laser power of 100 mW (a), and 60 mW (c). The red lines are the groove edges as determined with a MATLAB edge-detection routine. The groove widths were measured at regular intervals, and histograms of the groove width distribution are shown: panel (b) is the histogram for the 100-mW groove in (a), and panel (d) for the 60-mW groove in (c). Black curves are Gaussian fits, with a full-width at half-maximum of 4.3 μm (b) and 2.4 μm (d). We ascribe the improved smoothness of (c) compared to (a) to the lowering of the laser power used in the etching process.

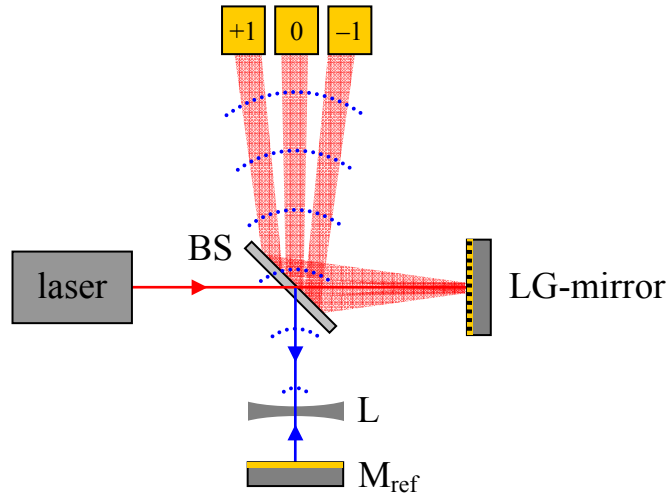

**Fig. 9.** Schematic of Michelson interferometer used to observe intensity and interference patterns of optical vortices produced by laser-etched LG-mirrors (BS is a beam splitter). A removable diverging lens, L, was used to create the spherical reference beam. The vortex beam is drawn in red, and the reference beam in blue. By removing the diverging lens a plane reference wave can be interfered with the vortex beam. We investigated the diffraction orders +1, 0, and −1 (yellow boxes).

The setup was radiated with full laser power (~2.3 W). The reference arm was blocked with an opaque screen allowing only the vortex beam to emerge from the interferometer. The resulting vortex beam was sent through a 1-m focusing lens and images were observed in the focus using a CCD camera. We placed neutral-density filters in front of the camera to avoid damage. Images of vortices of charge 1, 2, and 7 are shown in Figs. 10(a,b,c). Removing the opaque screen allowed the reference beam to pass through the interferometer with a planar wavefront. Interference with this wavefront resulted in an intensity pattern which mimics the LG mirror lines, seen in Figs. 10(d,e,f). Finally, by adding the diverging lens to the reference arm (as shown in Fig. 9), a spherical reference beam was allowed to interfere with the vortex beam, resulting in the spiral structures seen in Figs. 10(g,h,i).

The observed vortices in Figs. 10(a,b,c) show nonzero intensity at their centers. This is due to angular dispersion resulting from the diffraction of broadband (~20 nm) radiation[18]. To compensate this angular dispersion we used a folded version of our $2f$-$2f$ setup[15,16]. In this folded setup, Fig. 11, the radiation passes through the converging lens L1 twice and there is a negligible distance (~150 μm) between the lens L1 and the folding mirror M. Therefore, if we mentally unfold the setup the effective focal length, $f_{\text{eff}}$, follows from

$$\frac{1}{f_{\text{eff}}} = \underbrace{\frac{1}{f}}_{\text{first pass}} + \underbrace{\frac{1}{f}}_{\text{second pass}} = \frac{2}{f} \quad \Rightarrow \quad f_{\text{eff}} = \frac{1}{2}f = 50 \text{ cm} \qquad (9)$$

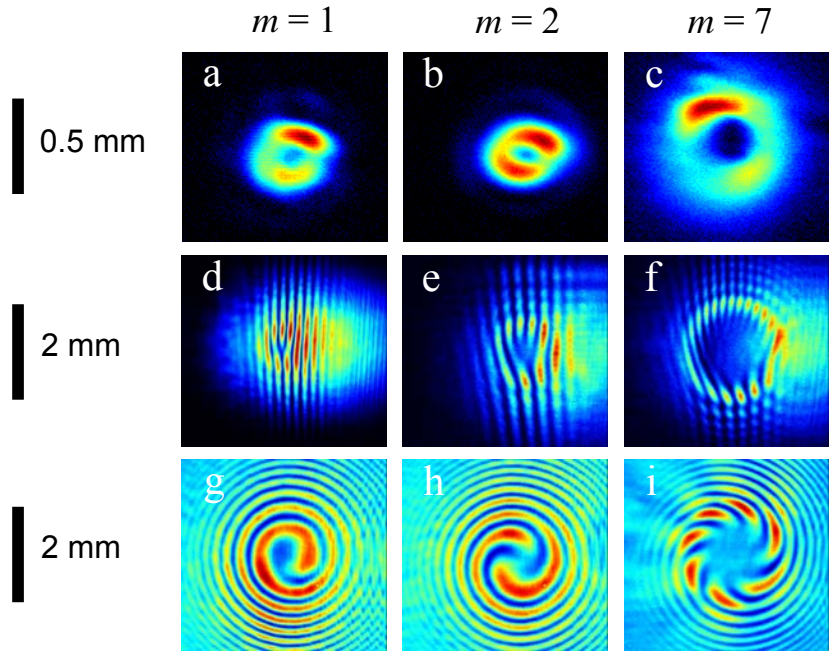

**Fig. 10.** Intensity profiles and interference patterns for optical vortices with charge 1, 2, and 7 (left, center, and right columns). First row (a,b,c): far-field images of optical vortices taken in the focus of a 1 m lens. Second row (d,e,f): interferograms of the optical vortices with a plane reference wave; these images mimic the LG-grating pattern. Third row (g,h,i): images of the optical vortices interfering with a spherical reference beam, creating a spiral intensity pattern. The vortex charge is confirmed by counting the number of intertwined spiral arms in these interferograms.

and so we adjusted the distance between the LG mirror and L1/M to 100 cm = $2f_{\text{eff}}$. When compensating the angular dispersion, we used our Spectra-Physics Tsunami oscillator, whose bandwidth (~45 nm) is about twice that of the amplified beam (~20 nm). Pulse durations for the oscillator were determined to be ~110 fs from frequency-resolved optical gating (FROG) measurements. This pulse duration is not transform-limited due to pulse stretching by the final cavity optics. (An external compressor would compress the pulse.) The resulting ±1 diffraction orders emerging from the folded $2f$-$2f$ setup show compensated, Figs. 12(a,b,c), and uncompensated, Figs. 12(d,e,f), vortex profiles. The uncompensated −1 order has twice the angular chirp that it would have had if it had been diffracted by only one grating[15]. Finally, a pinhole was placed in the zero order beam to create a spherical reference beam. The resulting interferograms, shown in Figs. 12(g,h,i), confirm the topological charge we etched.

To determine the maximum intensity that the LG-mirrors can withstand before becoming irreversibly damaged, an LG-mirror was placed on an optical track so that it could be moved along the optical axis of a 1 m focusing lens. The lens focused 800 nm radiation from our Ti:sapphire laser amplifier having a pulse duration of around 50 fs. At a laser intensity of about $10^{12}$ W/cm$^2$, self-focusing was observed in the quartz glass behind the reflective surface of the LG-grating. The intensity was increased by moving the LG-mirror further into the focus. It was observed that the reflective gold coating became damaged at intensities >$10^{12}$ W/cm$^2$. This value is in agreement with [22]. To ensure that the LG mirror could withstand these intensities for long periods of time, the mirror was left for 30 minutes at slightly less than $10^{12}$ W/cm$^2$. Microscope observations revealed no damage to the LG-mirror after this exposure.

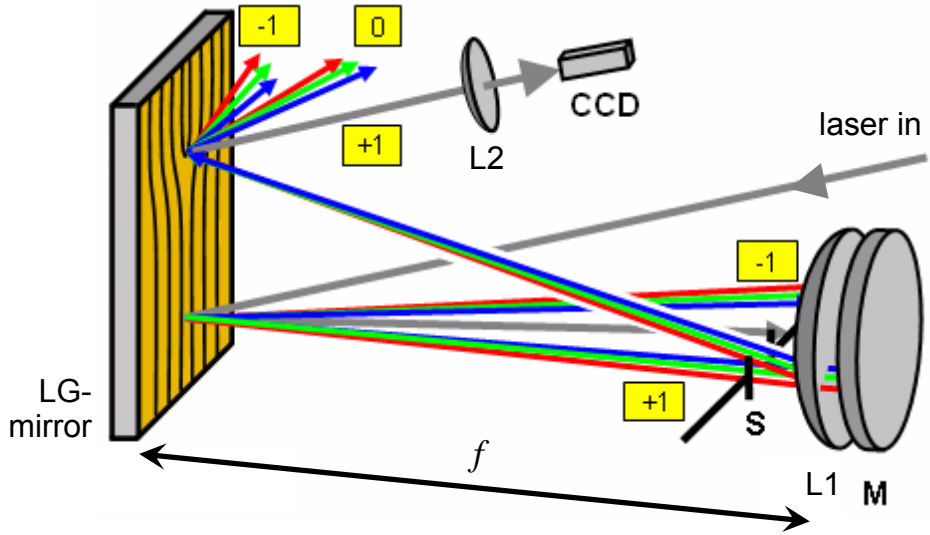

**Fig. 11.** Schematic of the folded $2f$ - $2f$ setup (angles, object dimensions and relative object positions not to scale for clarity). Ultrashort laser pulses enter from the top right, and then propagate along the following path: bottom half of LG-mirror, containing the line grating of which a part is shown in Fig. 2(a); S, order-selecting aperture, blocking all diffraction orders but +1; L1, plano-convex lens with focal length $f$ = 100 cm; M, folding mirror; top half of LG-mirror, containing a grating with the vortex fingerprint of which an example appears in Fig. 2(b); L2 is a collimating lens; CCD is a CCD camera. The dimensions and relative positions of the optical elements are as follows: the etched part of the LG-mirror is square, 2 cm × 2 cm; distance between LG-mirror and L1 is 100 cm; distance between L1 and M is ~150 μm; distance between LG-mirror and L2 is 25 cm; CCD is located 120 cm behind L2. Diffraction orders (−1, 0, +1) are indicated as black numbers on a yellow background. A few colored rays are shown to remind the reader that there is spatial chirp, with gray rays indicating that spatial chirp is absent. Note that we show just three colors—in reality, the frequency spectrum is of course continuous.

Inspection of the images in Fig. 12 shows that the LG-mirrors produce vortices of the correct topological charge. Also, the angular chirp they produce can be compensated using existing methods. Thus, we made fully functional vortex-producing gratings based on the skeleton equations we presented in Sec. 2. The scattered light seen in Fig. 12 must be at least in part due to the non-perfect groove smoothness of our laser-etched holograms. This background noise causes the vortex of charge 2 that was expected in Fig. 12(b) to split into two adjacent vortices of charge 1[6]. It also leads to a nonzero intensity in the heart of the vortex of charge 7 in Fig. 12(c). Improvement of the smoothness is expected to improve the quality of the vortices we produce.

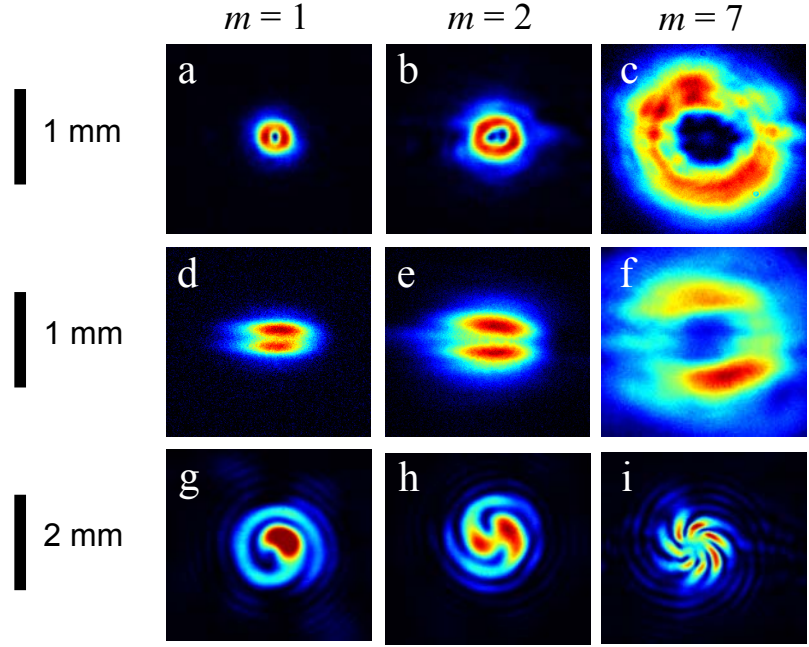

**Fig. 12.** Intensity profiles and interference patterns for optical vortices with topological charges 1, 2, and 7 (left, center, and right columns) after compensation of angular chirp using the folded 2*f*-2*f* setup. First row (a,b,c): far-field images of the +1 order (compensated) optical vortices. Second row (d,e,f): far-field images of the −1 order (uncompensated) optical vortices. Third row (g,h,i): images of the compensated optical vortices in the focus, interfering with a spherical reference wave. This wave was created by placing a pinhole in the zero order beam. As in Fig. 10, the vortex charge is confirmed by counting the number of intertwined spiral arms in the interferograms.

**5. Conclusions and Outlook**

In conclusion, we have demonstrated a simple and straightforward way to produce optical vortices by laser-etching grating lines into typical laser-quality gold mirrors. We have shown that these LG-mirrors are sufficiently smooth and withstand high intensities. Their gold plating also allows for large bandwidths, making them suitable for a broad range of applications. Future experiments will involve laser-etching silver and dielectric mirrors. In addition, improvement can be made to increase the efficiency of these LG-mirrors and/or setups in which they are used when producing femtosecond optical vortices. First, it has been shown that small misalignments to a compressor allows precompensation of angular dispersion[16]. The elimination of the first grating pass in the 2*f*-2*f* setup increased the efficiency by an order of magnitude when binary gratings were used. Second, binary blazing techniques[23] might improve the first-order diffraction efficiency by an additional factor of four. With these techniques, laser-etched LG-mirrors would remain suitable for large bandwidth pulses in a wide range of optical wavelengths while retaining much of the power in the compensated order.

**Acknowledgments**

This material is based upon work supported by the National Science Foundation under Grant No. PHY-0355235. T.S. acknowledges an REU grant from the National Science Foundation.